\documentclass[
    ,final            
  ]
  {aipproc}

\layoutstyle{8x11single}

\usepackage{epsfig}
\usepackage{rotating}

\def        \dzero  {D\O~}

\begin{document}

\title{Search for Single Top Quark Production\\Using Likelihood Discriminants\\at \dzero in Run~II}


\author{ Shabnam Jabeen \\
           for the \dzero Collaboration}{
}

\begin{abstract}
We present an improved search for single top
quarks in two production modes, $s$-channel ($tb$) and $t$-channel
($tqb$). The search is performed in the electron+jets and muon+jets
decay channels, with one or more $b$-tagged jets, on nearly
370~pb$^{-1}$ of
\dzero Run~II data collected between
August 2002 and October 2004. Impact-parameter based b-quark tagging is used to select signal-like events.
We use a likelihood discriminant method to separate signals from
backgrounds. The resulting expected/observed $95\%$
confidence level upper limits on the single top quark production cross
sections are 3.3/5.0~pb ($s$-channel) and 4.3/4.4~pb ($t$-channel).

\end{abstract}

\maketitle


\section{Introduction}
\vspace{-0.15cm}
The top quark was originally discovered in 1995, at the Fermilab Tevatron $p{\bar p}$ Collider Run I by the CDF and
\dzero collaborations~\cite{topdiscovery}. It was observed in its $t\bar{t}$ production mode via the strong interaction ($q\bar{q}\rightarrow g \rightarrow t\bar{t}$).\\
Within the Standard Model, another production mode via the electroweak interaction is possible. This mode is
called single top quark production as only one top quark is produced with
another $b$ quark through the $Wtb$ vertex. As a consequence, a measurement of the single top quark production cross section can
be used to constraint the magnitude of the CKM matrix element $V_{tb}$ and study the properties of the $Wtb$
coupling. The two
main Feynman diagrams for $s$- and $t$-channel single top quark production at the Tevatron Run~II are given in Fig.~1.\\

Single top quark production has not yet been observed  and is more challenging than $t\bar{t}$
production due to smaller cross sections (2.86~pb in total, with $s$-channel cross section to be $0.88 \pm 0.14$~pb and $t$-channel cross section to be  $1.98 \pm 0.30$~pb) and a much larger, less
discriminable background.
We present a new analysis of $\sim370$~pb$^{-1}$ of \dzero Run~II data using a likelihood discriminant
method to separate signals and backgrounds and we derive 95\% confidence level upper limits to the $s$- and $t$-channel
single top quark production cross sections.
\begin{figure}[!h!tbp]
  
      \includegraphics[width=0.8\textwidth]{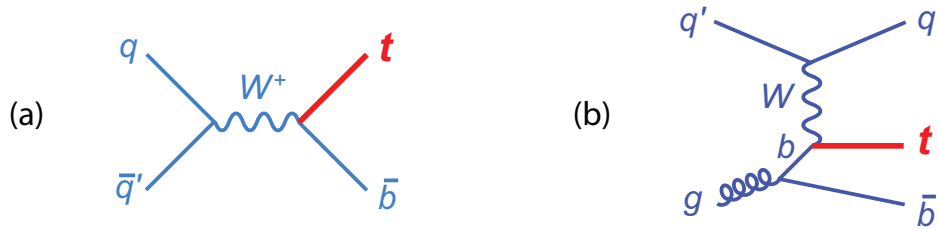}
  
\caption{The dominant Feynman diagrams for
single top quark production at the Tevatron $p{\bar p}$ Collider:
$s$-channel, $tb$ final state (left diagram) and $t$-channel,  $tqb$ final state (right diagram). In this note, we use the simplified notation $tb$ and $tqb$ which
implicitly includes all possible charge conjugations. }
\end{figure}

\section{The D0 detector}

The \dzero detector for Run~II, completely described in~\cite{D0detector}, consists of a central tracking system, a liquid-argon/uranium sampling calorimeter
and an iron toroid muon spectrometer. The central tracking system is composed of a silicon microstrip tracker (SMT) and a central fiber
tracker (CFT), both located into a 2T superconducting solenoidal magnet. The SMT  detector has about 800000 individual strips and its design is
optimized for tracking and vertexing capabilities allowing heavy flavour tagging. The calorimeter is longitudinally segmented into
electromagnetic and hadronic layers and is housed into three cryostats: a central barrel covering $|\eta|\leq 1.1$ and two
end-caps that extend coverage up to $|\eta|\leq 4$. The muon system resides beyond the calorimeter and
consists of  a layer of tracking detectors and scintillation counters before the toroidal magnet, followed by two similar layers after the
toroid. Tracking in the muon system relies on wide or mini drift tubes depending on the acceptance (up to $|\eta|=2$).

\section{Analysis Overview}

This analysis focuses on the final state
topology where the top quark decays into a $b$ quark and a $W$ boson, which
subsequently decays leptonically ($W\rightarrow e\nu ,\mu\nu$). 
 The general selection 
is designed to reject misreconstructed events and to select a
signal-like data sample that is well reproduced by Monte Carlo backgrounds samples. The $b$-tagging requirement enhances the discrimination between signals and the $W$+light jet and multijet backgrounds.
We estimate, mostly from Monte Carlo samples, the yields for the main backgrounds after the final procedure
which includes  trigger and  selection effects as well as
$b$ quark jet tagging.
Due to the presence of two dominant classes of background with different kinematic properties ($W$+jets and $t{\bar t}$-like events), two likelihood discriminant variables are built. The estimated yields and likelihood
distributions for the backgrounds are
confronted to the number of observed events in data to extract 95\% confidence level upper limits using a
Bayesian fit.

\section{Likelihood discriminant method}
\vspace{-0.15cm}
After the event selection, a final discriminating variable is constructed in
order to efficiently characterize the signal type events and reject the background
type ones, based on the shapes of the mostly uncorrelated input variables.

 Examples of likelihood filters outputs for signal and backgrounds are given in Fig.~2 and  Fig.~3.

\begin{figure}[!h!tbp]
\includegraphics[width=0.3\textwidth]{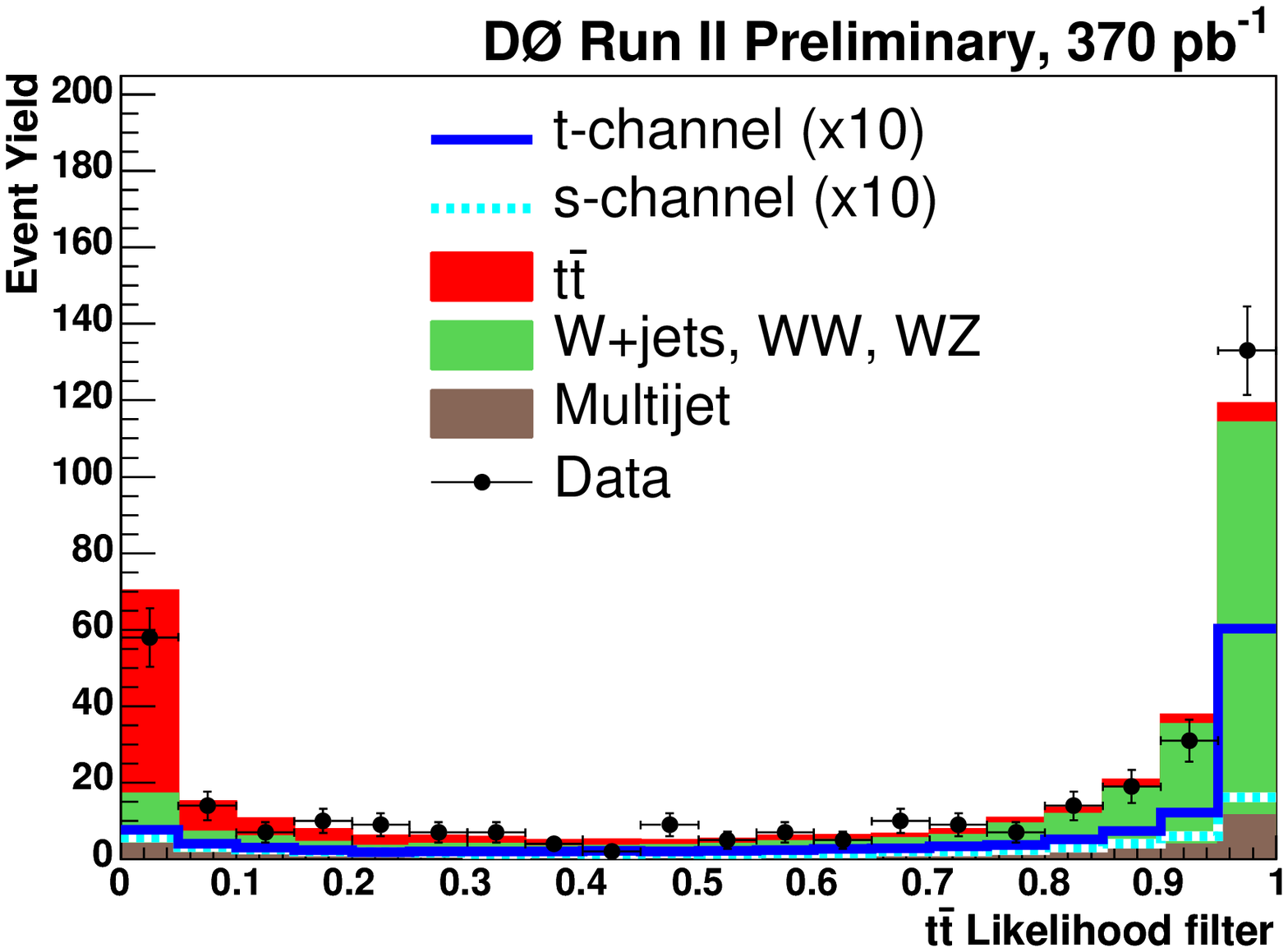}
\hspace{.3in}
\includegraphics[width=0.3\textwidth]{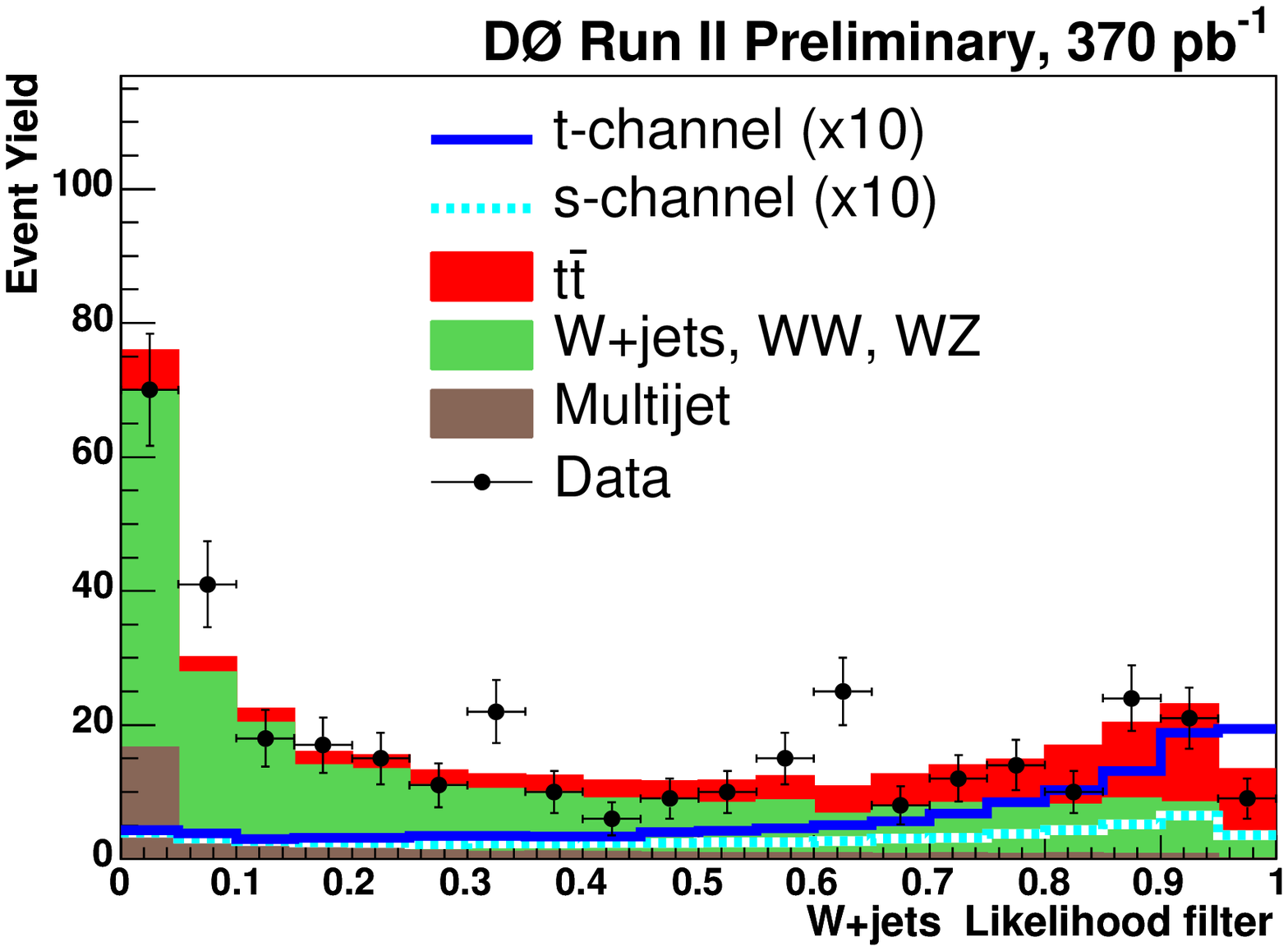}
\caption{Electron and muon channels combined. Data to Monte-Carlo comparison for the $tqb$/$t\bar{t}$ (left) and $tqb$/$W$+jets (right) for single tagged events.}
\end{figure}

\begin{figure}[!h!tbp]
\includegraphics[width=0.3\textwidth]{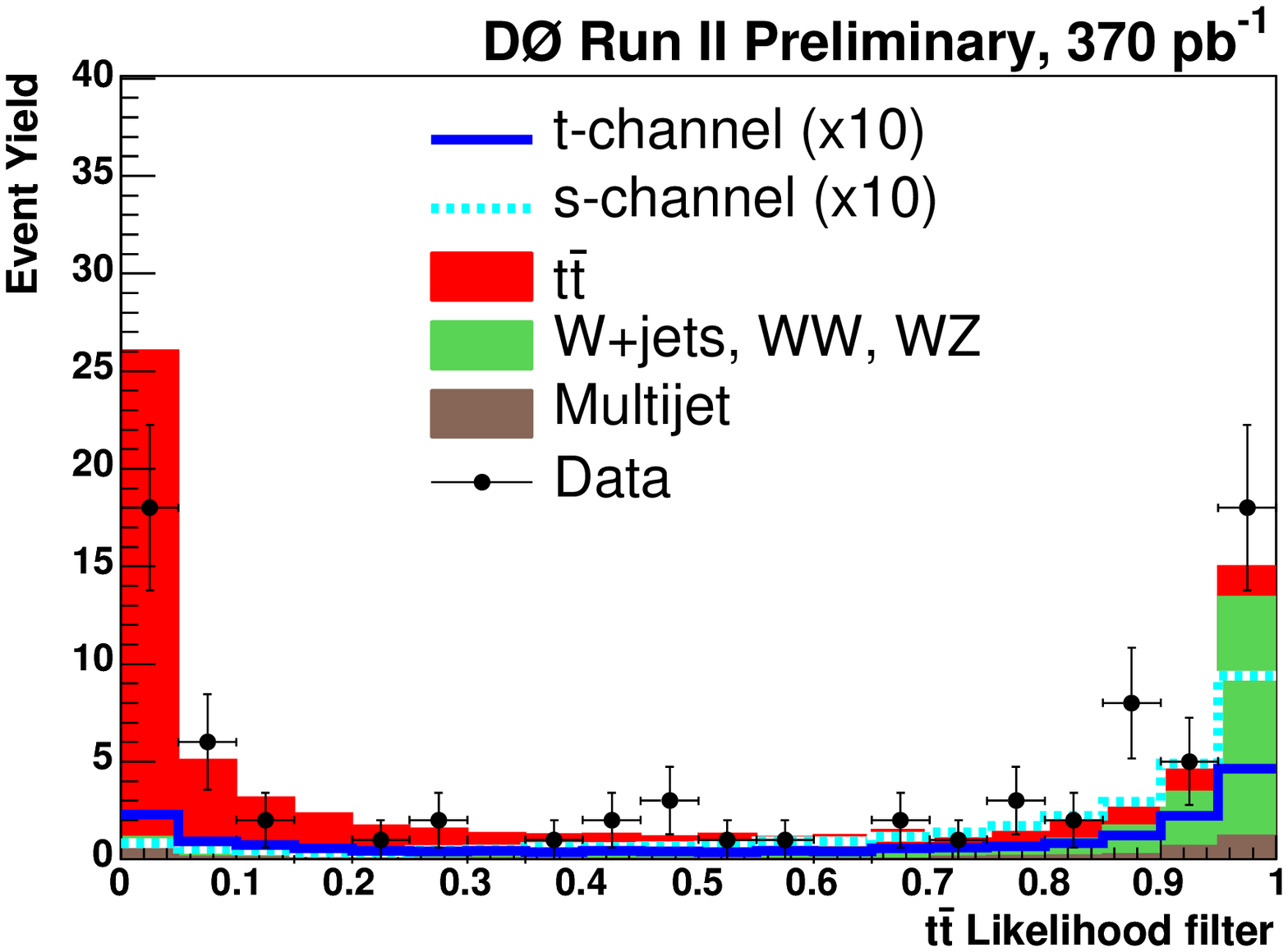}
\hspace{.3in}
\includegraphics[width=0.3\textwidth]{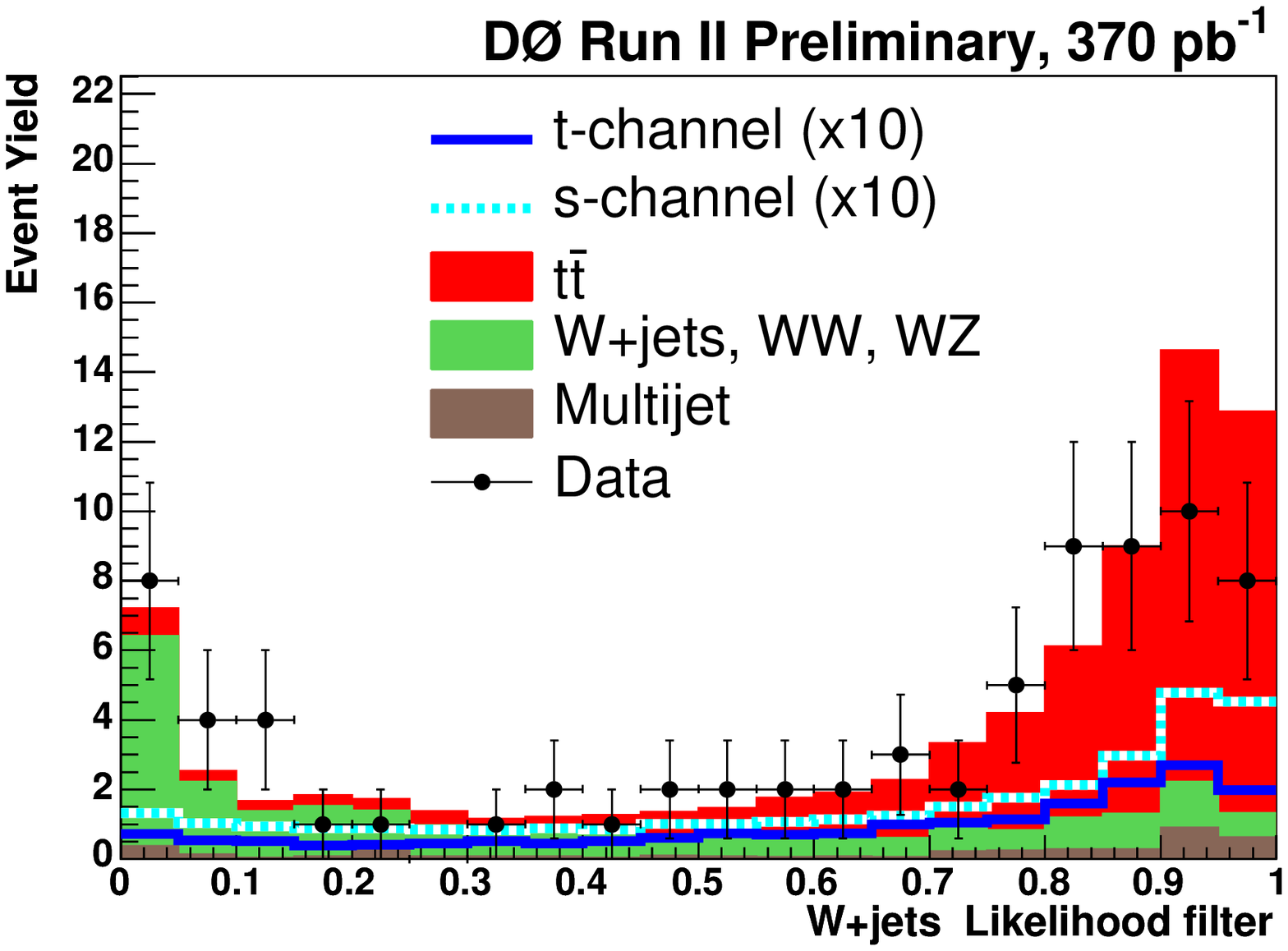}
\caption{Electron and muon channels combined. Data to Monte-Carlo comparison for the $tb$/$t\bar{t}$ (left) and $tb$/$W$+jets (right) for double tagged events.}

\end{figure}

\section{Cross section limits}

The number of observed events is consistent with the background prediction for both muon and electron channels and for all $b$-tagging schemes, within the total uncertainties. We therefore set upper limits at the 95\% confidence level, using a Bayesian approach~\cite{IainTM2000}.

The observed (expected) 95 \% confidence level limits are 5.0~pb (3.3~pb) for the $s$-channel and 4.4~pb (4.3~pb) for the $t$-channel.
\begin{figure}[!h!tbp]
\includegraphics[width=0.45\textwidth]{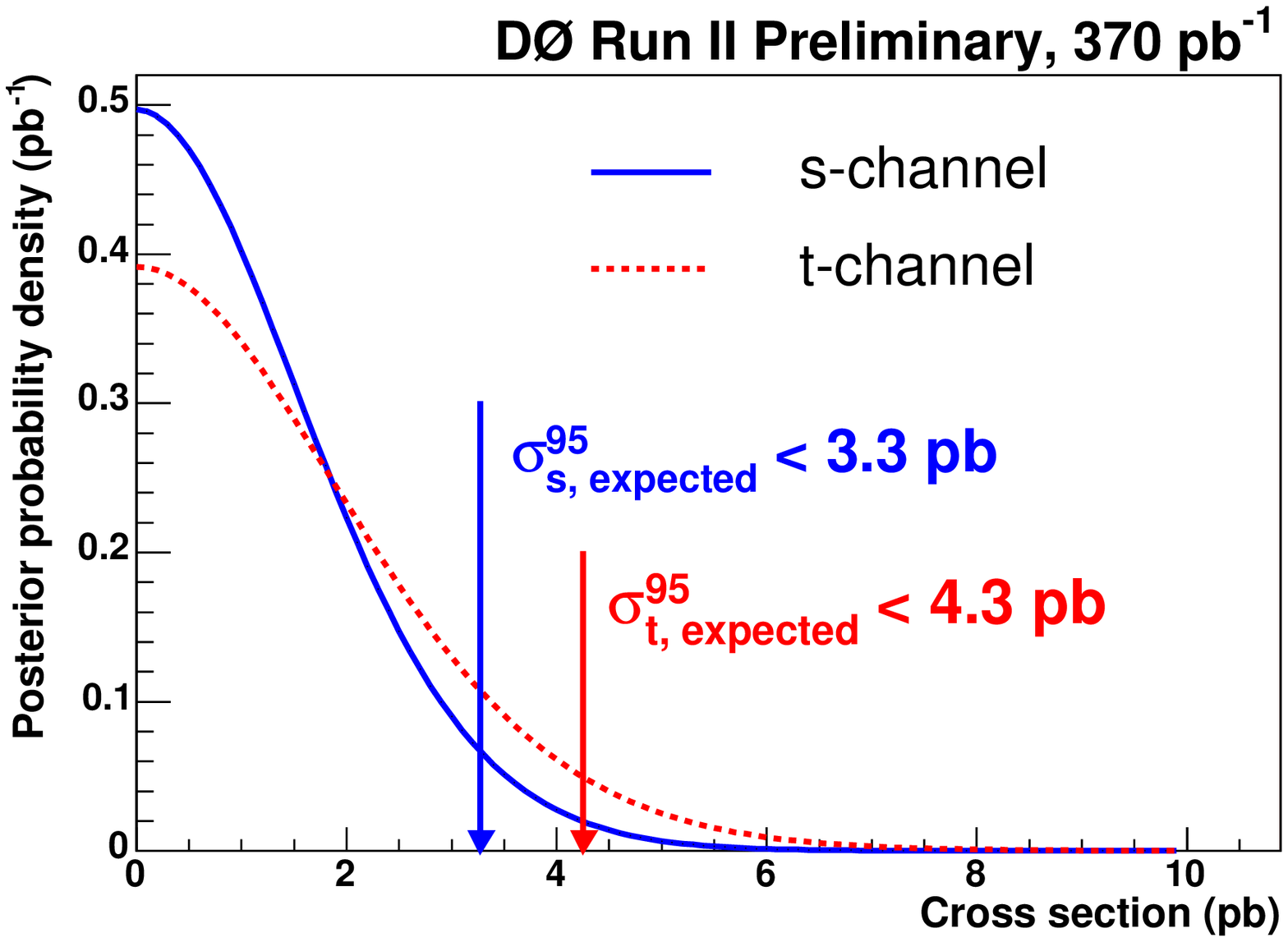}
\hspace{.1in}
\includegraphics[width=0.45\textwidth]{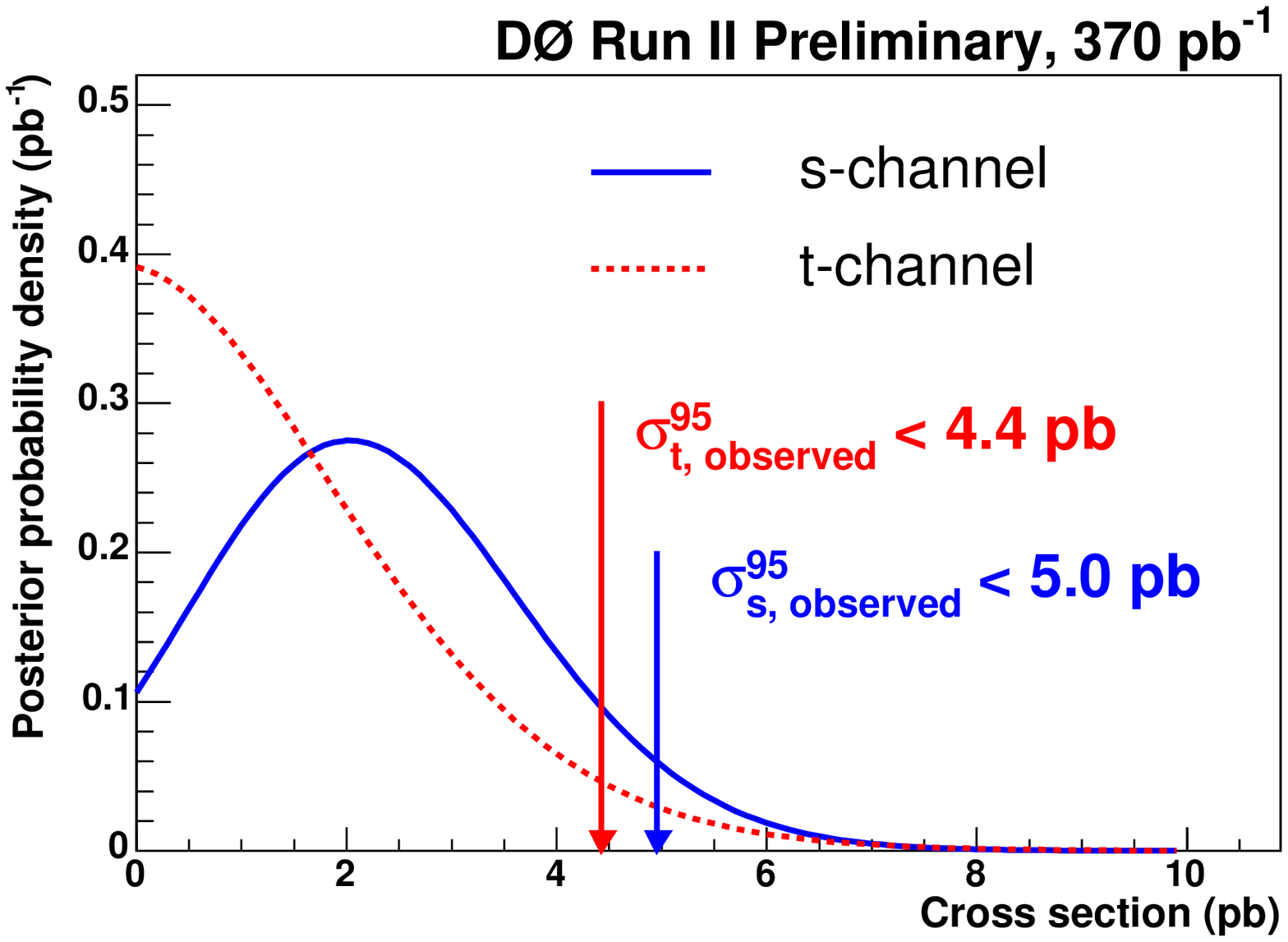}
\caption{Expected (left) and observed (right) Bayesian posterior densities with 95\% confidence level limits for combined electron and muon channel and combined $b$-tagging schemes.}
\label{fig-post}

\end{figure}

This result improves the previous limit published by the \dzero Collaboration~\cite{RunII:d0_result}. Both analyses have very similar strategies, the main differences being the b-tagging algorithm, the final discriminant and the integrated luminosity.  The sensitivities of both methods are very similar.

\section{Summary}
\vspace{-0.15cm}
We analyzed nearly 370 pb$^{-1}$ of data collected by the \dzero Run~II detector. No evidence has been found for electroweak production of the top quark in either $s$- or
$t$-channel. Separate analyses in electron+jets and muon+jets final state, with one or two identified $b$ quark jets were combined to improve the sensitivity. Upper limits
at the 95\% confidence level on the cross section for each $s$- and $t$-channel processes have been set using a Bayesian fit to likelihood discriminant distributions.
The final limits  for the $s$-channel and for the $t$-channel are found to be of 5.0~pb, and  4.4~pb, respectively.







\begin{thebibliography}{99}

\bibitem{topdiscovery}
F.~Abe {\it et al.}, (CDF Collaboration),
``Observation of Top Quark Production in $p \bar{p}$ Collisions,''
Phys.\ Rev.\ Lett.\ {\bf 74}, 2626 (1995);
S.~Abachi {\it et al.}, (\dzero Collaboration),
``Observation of the Top Quark,''
Phys.\ Rev.\ Lett.\  {\bf 74}, 2632 (1995).



\bibitem{RunII:d0_result}
V.M.~Abazov {\it et al.}, (\dzero Collaboration), ``Search for single top quark production in pbarp collisions at sqrt(s)=1.96 TeV''
hep-ex/0505063; Fermilab-Pub-05/207-E, submitted to Phys. Lett. B (2005).


\bibitem{D0detector}
V. Abazov {\it et al.}, D{\O} Collaboration, in preparation for
submission to Nucl. Instrum. Methods in Phys. Res. {\bf A};
T. LeCompte and H.T. Diehl, Ann. Rev. Nucl. Part. Sci. {\bf 50},
71 (2000).


\bibitem{IainTM2000}
I. Bertram et al., ``A Recipe for the Construction of Confidence
Limits,'' Fermilab-TM-2104 (2000).
\end{thebibliography}

\IfFileExists{\jobname.bbl}{}
 {\typeout{}
  \typeout{******************************************}
  \typeout{** Please run "bibtex \jobname" to optain}
  \typeout{** the bibliography and then re-run LaTeX}
  \typeout{** twice to fix the references!}
  \typeout{******************************************}
  \typeout{}
 }



\end{document}